# Influence of Spinning Electric Fields on Natural Background γ-Radiation

**Mark Krinker**
*The Member of Advisory Committee of City College of Technology, Department of Electrical Engineering & Telecommunications, CUNY, New York*
mkrinker@aol.com

**Felix Kitaichik,**
*Engineer, Key Systems, New York*
kitai4ik@yahoo.com

**Abstract:** This paper considers influence of spinning electric field on statistics of natural background gamma-radiation. The spinning electric field, shown as a virtual gyroscope, has quantum mechanics characteristics. Interaction of the virtual fermion-like gyroscope with bosons (gamma-quanta) results in lowering intensity of the gamma-radiation and altering Poisson distribution. The statistic of the observed phenomenon depends on the direction of rotation of the virtual gyroscope. The results are discussed in a shade of spin-spin interaction having regard to realizing thermodynamically profitable conditions. Similarity of observed reduction of gamma-radiation in spinning electric fields and that for mechanical rotation stresses a special role of rotation itself, disregarding the matter of its carrier.

## 1. Introduction.

Rotation plays a special role in a paradigm of modern physics. Historically, rotation was considered as a base of matter by Rene Descartes back in XVII century and even before in Eastern Philosophies. On a modern level it descends to the conception of torsion fields introduced by A. Einstein and E. Cartan in early XX century, but only in the late of the century real impetus was given to it. On the phenomenal level, the special role of rotation first was observed by Russian Prof. N.P. Myshkin in the experiments with scales back in the late XIX century. [1]

The special role of rotation was then shown by Soviet physicists N.A Kozyrev in his theoretical and experimental works with gyroscopes [2] and later by A.I.Veinik in his study of chrono-related effects [3].
It's important to stress that Kozyrev and his colleagues observed variation of the weight of gyroscopes with non-stationary spinning. The results of Kozyrev's experiments on the high-speed non-stationary gyroscopes were confirmed by other researchers [4].

The theory and practice of the torsion fields got a powerful development in works of G.I Shipov and A.E. Akimov in 90's and our days. The good collection of the related works of these and other authors we find in [5].

The researches of our days revealed new experimental facts, accompanying rotational motion, which are beyond the scope of the classical concepts. These studies can be subdivided for several groups.

In one of them, dynamic effects of non-stationary mechanical rotations are studied. The author of [6-8] studied unbalanced rotation of parallel disk and found origination of attraction force between them.

Another group of studies revealed influence of mechanical rotation on statistics of the gamma-decay. First reports on that, Tomsk, 1995, claimed that mechanical rotation breaks the shape of Poison distribution [9]. Later profound research [10] showed reducing gamma-decay under influence of remote rotation as well as distortion of normal shape of the Poisson distribution. It was proved that mechanical rotation influences spectrum of gamma-radiation of $Cs^{137}$ $Co^6$.

Having spinning wheels to cause the effect of reduction of gamma-radiation isn't only mechanical driver of the effect. For instance, the authors of the recent publication [11] state that exploitation of vortex-base heat generators (the torsion technology) reduces background radiation from 20-12 uRm/h to 10-6 uRem/h.

The third group is a rotation-stimulated space-time distortions. Profound impact of rotation on flowing time was studied in [12]. The authors found unequal action of clockwise and counterclockwise rotation, and what's more, the clockwise rotation causes more profound action. It's important to stress here, that the authors employed non-stationary rotation achieved with additional high-speed spinning disc installed in vicinity of the major slow rotating object.

Looks like that most exotic of all the torsion effect sources are the passive generators on a base of shape-effect. The authors of [12] report reduction of the background gamma-radiation from 14 to 8 uRem/h in vicinity of special geometric forms.

All the considered above studies have a common denominator in that they are manifestations of torsion/ spinning field phenomena. What observed is a secondary and depends on a type of the detector we use. Majority of the considered studies are based on mechanical rotations-caused phenomena.

The authors of the present study consider electric field rotation as a source of the similar phenomena. The purpose of the current research is proving equivalency of electrical and mechanical rotation in terms of torsion approach.

Earlier, the conception of the spinning electric field as a Virtual Gyroscope was developed by the author [13,14]. This approach was a logical extension of well-known studies of N.A. Kozyrev on a special role of spinning motion. Kozyrev studied phenomena of high-speed mechanical gyroscopes and found variation of their weight as they spun. The results of Kozyrev's experiments on the high-speed gyroscopes were confirmed by other researchers. It has to be said, that the successful experiments were based on non-stationary rotation [4].

Conception of the Virtual Gyroscope considers spinning of electrical field and its angular momentum as a real analog of mechanical rotation that already proved its influence of fundamental physical processes.

The purpose of the current research is proving equivalency of electrical and mechanical rotation in terms of the torsion approach.

It was proved that mechanical rotation influences spectrum of gamma-radiation of $Cs^{137}$ $Co^{6.}$

If the fact of rotation is more important than a carrier of the process, then we could expect that the Virtual Gyroscope can show similar influence on statistics of the radioactive decay.

Our previous experiments on interacting non-electromagnetic spinning axion fields, induced by the generators, developed by A.A. Shpilman [15], revealed interaction between these fields and the spinning electric ones [16]. And what's more, the remaining electric spinning was observed after turning off the axion generators. This remaining rotation fits in our conception of the Virtual Gyroscope

First, let's us briefly recall the conception of the Virtual Gyroscope in a quantum paradigm.

## 2. Virtual Gyroscope

### 2.1. General Approach

Virtual gyroscope is a spinning localized field, having a certain center of mass and angular momentum. Unlike elliptically polarized electromagnetic waves, energy remains localized, that allows considering its mass behaving like one of a real gyroscope.
It has a moment of inertia

$$I = \frac{mr^2}{2} \tag{1}$$

Its equivalent mass can be derived from the sum of its time-varying potential and kinetic energies:

$$m = \frac{W}{c^2} = \frac{\varepsilon\varepsilon_0 \int_V E^2(t)dV + I(t)\omega^2}{2c^2} = \frac{\varepsilon\varepsilon_0 \int_V E(t)^2 dV}{2c^2 - 0.5r^2\omega^2}, \tag{2}$$

In this experiment, $\omega r \ll c$ and we can ignore it.

The spinning electric vector $E(t)$ of the virtual gyroscope is formed by two orthogonal (or having the orthogonal components) parent vectors $E_1(t)$ and $E_2(t)$, shifted in a phase for $\alpha$ degrees.

Electric rotation can be shown as a vector $S_e$, directed exactly as an angular momentum.

$$\vec{S}_e = \omega[\vec{E}_1(t)\vec{E}_2(t)] \tag{3}$$

### 2.2. Non-stationary spinning

According to N.A. Kozyrev, non-stationary rotation is a major condition for originating the weight-variation phenomena. We also see the non-stationary in some of the discussed above experiments.

Speaking of the rotating field, modulation is one of the ways of realizing the non-stationary conditions.

For two parent fields of equal frequency $\omega$, one of which is amplitude-modulated with the frequency $\omega_1$, we can write down

$$E_1(t) = E_{1m} \sin(\omega t + \varphi_1) \tag{4}$$

$$E_2(t) = E_{2m} \sin(\omega t + \varphi_2)\sin(\omega_1 t + \varphi_3), \tag{5}$$

that is

$$E_2(t) = \frac{1}{2} E_{2m} [\cos((\omega - \omega_1)t - (\varphi_2 + \varphi_3)) - \cos((\omega + \omega_1)t + (\varphi_2 + \varphi_3))] \tag{6}$$

We see origination of combined frequencies in the non-stationary spinning.

Let's estimate the angular momentum of the virtual gyroscope.

Strictly speaking, we have to consider the total angular momentum of the virtual gyroscope as a sum of its spin and orbital portion

$$\vec{J} = \vec{L} + \vec{S}, \tag{7}$$

but its orbital portion is not developed as the spin one and we will not consider the orbital component.

Considering the field in the cells as a spinning cylinder of radius r, we have its angular momentum, which also can be shown in quantum way consisting of N elemental gyroscopes. Keeping a vector notation for the spin, we have

$$\vec{J} = \frac{mr^2}{2}\vec{\omega} \approx \frac{\varepsilon\varepsilon_0 \int_V E^2(t)dV}{4c^2} r^2 \vec{\omega} = \left\langle \hbar\sqrt{s(s+1)} \right\rangle N \tag{8}$$

Here, $E(t)$ is a spinning electric field.

The virtual gyroscope can be shown as a combination of N elemental quantum gyroscopes each having spin of 0.5 h/2π.

Basing on existence of the equivalent mass, spinning electric fields have to be considered as fermions rather than bosons. Fermions are usually associated with matter while bosons are believed to be force carriers.

It has to be said that the equivalent mass of the spinning field experiences periodic variation that can result in the slight variation of gravitation field.

## 3. The Experiment

**3.1. Installation**
The installation consists of two quadrupole capacitor cells, developing clockwise (left) and counterclockwise spinning (right) inside them (a top view). This allows using the cells as a differential analyzer in the various applications: as the shoulders of an interferometer, or placing identical matter in the cells having a differential thermocouple in the object and so on.

In the experiment, the electric spinning was realized with the electric vector, spinning at 39006 rad/s. The frequency is related to one of natural Earth-ionosphere resonator which, according to the authors vision, makes its contribution into originating so-called geo-pathogenic zones, GPZ. As it was earlier stated, the spinning electric field, driven by the fields of the Earth-ionosphere resonator, can result in GPZ-related phenomena including formation of torsion fields.

Thus, the experiment was based on creating two opposite spinning virtual gyroscopes. Each the gyroscope is a spinning 100 V/m electric field, rotating in a cell at 39006 rad/s. The experimental installation with the Geiger-Muller counter is shown in Fig.1.

The spinning inside the cells is a result of superposition of two orthogonal electric fields of equal frequency, shifted in phase for 45 degrees. The parent fields are induced by the plates of a quadrupole capacitor which is also an experimental cell. One of the principal features of the installation is that the spinning field is modulated. **This is nothing but realization on a field level requirement of N.A. Kozyrev on necessity of superimposing non-stationary conditions on mechanical rotation in his experiments with the gyroscopes.**

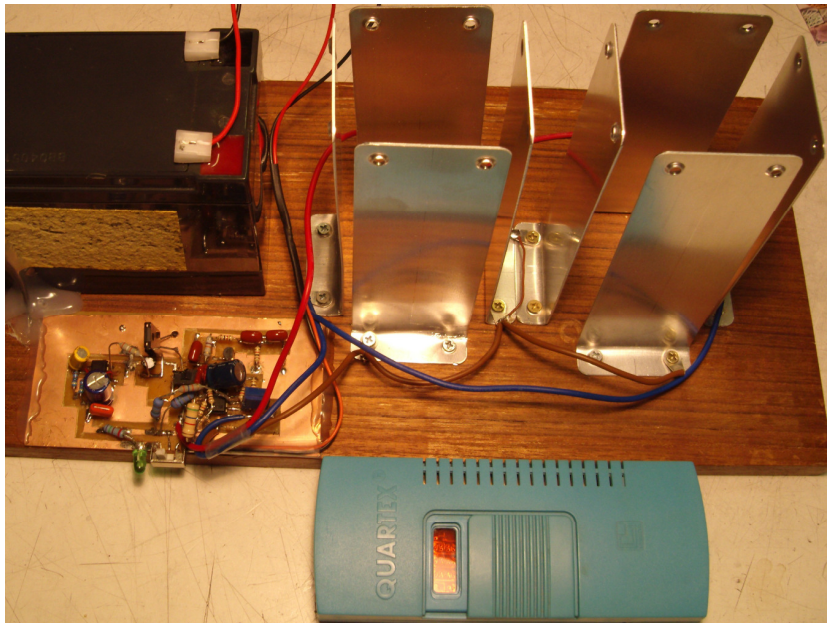

Fig.1
Installation for differential analysis of electric spinning.
The Analyzer also allows carry on the experiments on interference probing space in cells and other comparative measurements.

Fig.2a shows electric spinning with no modulation, while 2b shows the modulated spinning. The modulation was carried on 30 Hz as amplitude alternating of one of the parent fields plus some non-linear distortions.

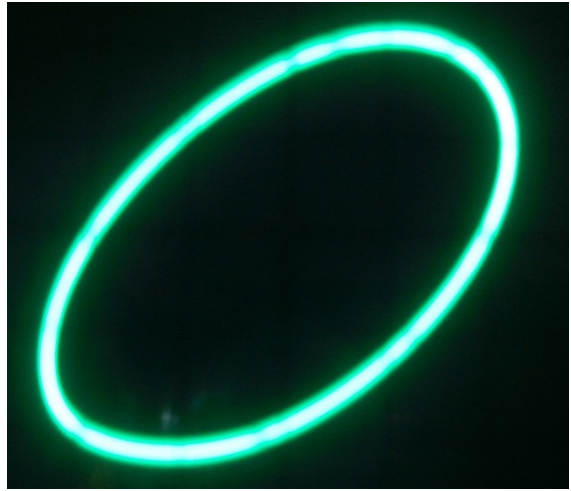

Fig.2a
Basic spinning the vector in the cells

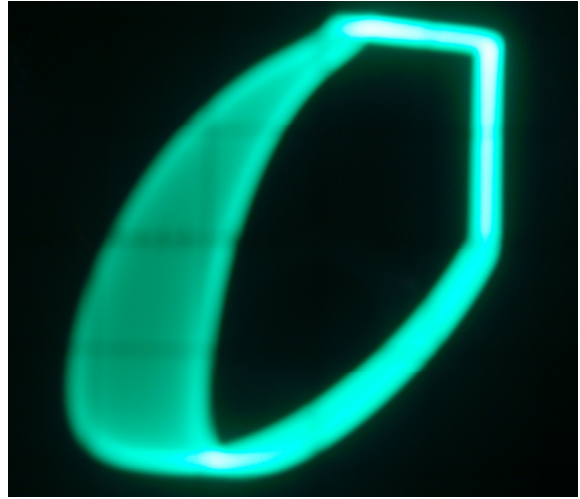

Fig.2b
Non-stationary rotation of the electric vector in the cells

The Geiger-Muller counter Quartex [17] was placed inside the cells. The counter detects X, Gamma rays and beta particles (X, β and γ). Its measurement cycle is 38 seconds. It measures radiation in micro Rem per hour (μRem/h)

Energy of X and γ quanta detected : 100 keV to 1,2 MeV, for β particles detected : 350 keV to 1,2 MeV.

Each the cycle of the measurement includes a reference sampling of 10 counts for 38 s, then the counter was placed inside a clockwise spinning cell with the following 10 counts and then moved into counterclockwise cell for the next 10 counts. There were 12 repetitions of the cycles, therefore the total number for each the group (reference, clockwise and counterclockwise) is 120.

## 3.2. The Results

The results are shown in the tables and the diagrams. Table 1 shows general statistics of the experiment, uRem/h, while Fig.3 shows it in a graphic way.

Table 1. Total statistics for each of the 120 counts in the cells for 38s sampling time.

|  | Reference | Counterclockwise | Clockwise |
|---|---|---|---|
| Average | 9.78 | 8.43 | 8.82 |
| Standard deviation | 3.16 | 3.09 | 2.42 |
| Square root of the average | 3.13 | 2.90 | 2.97 |

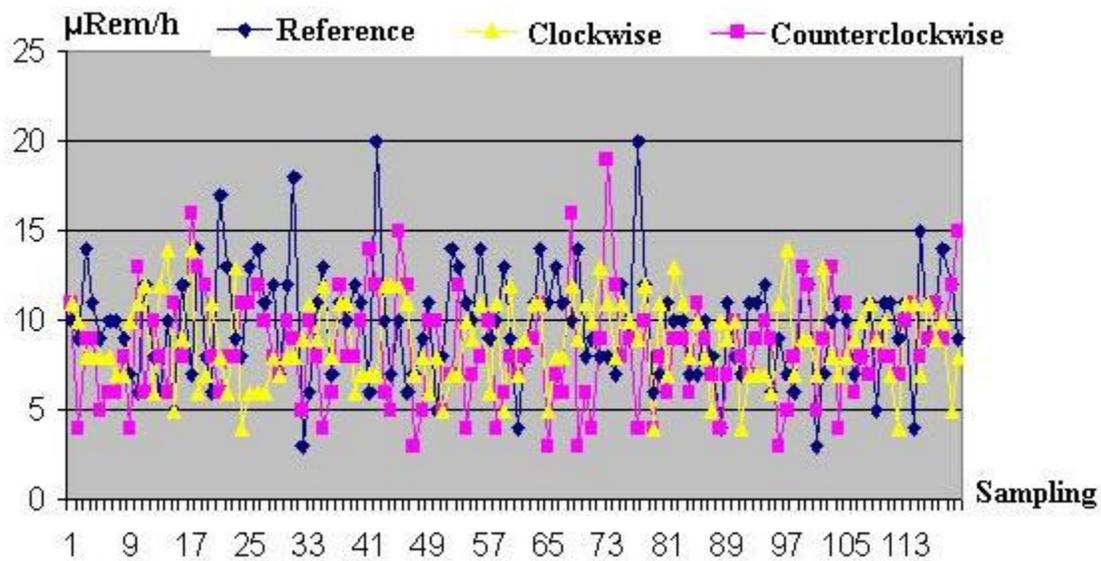

Fig.3

Total distribution of background radiation in the cells.

Table 2 shows the occurrence of equal counts in a row. The expected probability P of reading two equal numbers in a row in a series of A experiments equals $P = A\left(\dfrac{1}{N}\right)^n$, where N is a number of measurements in one cycle, n is a number of equal readings in a row.

Table 2. Occurrence of equal readings in the cells, having opposite spinning electric fields.

| Number of equal readings in a row | Reference | | Clockwise | | Counterclockwise | |
|---|---|---|---|---|---|---|
| | Experiment | Expected | Experiment | Expected | Experiment | Expected |
| 2 | 0.42 | 0.12 | 0.58 | 0.12 | 0.67 | 0.12 |
| 3 | 0 | 0.012 | 0.17 | 0.012 | 0 | 0.012 |
| 4 | 0 | 0.0012 | 0.08 | 0.0012 | 0 | 0.0012 |

Figs. 4-6 show examples of Poisson distributions for one cycle of the experiment.

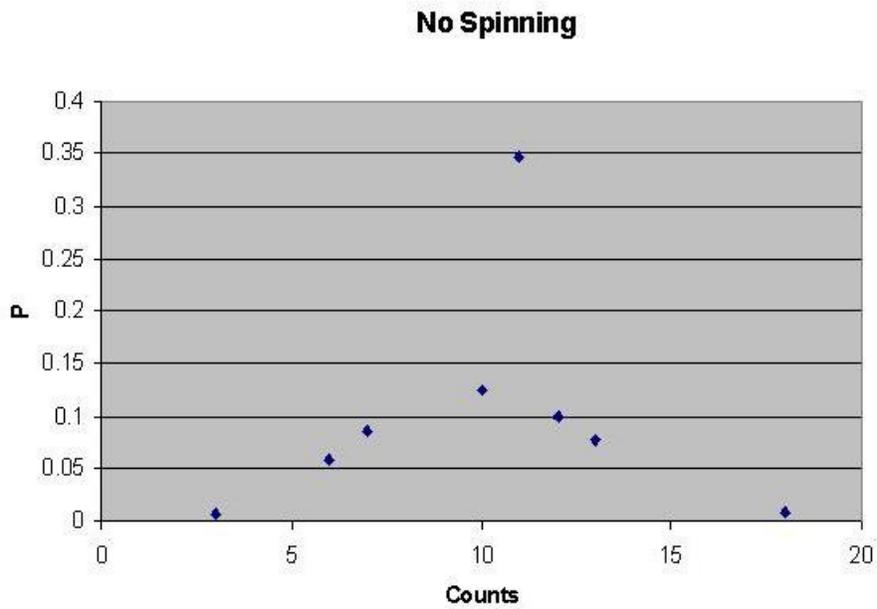

Fig.4

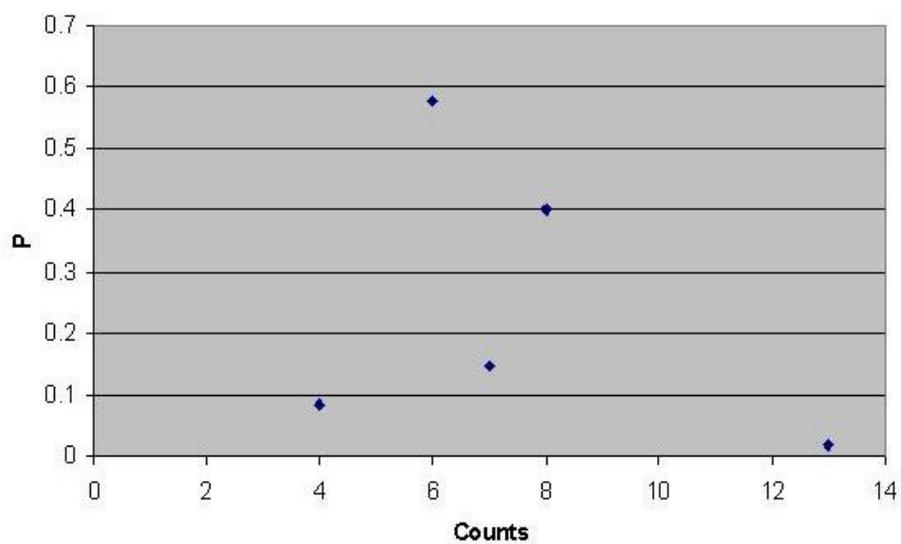

Fig.5

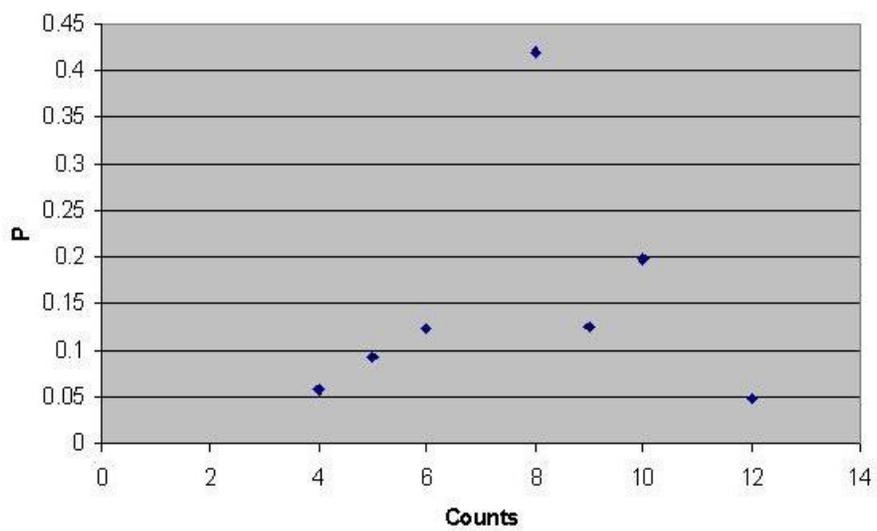

Fig.6

### 3.3. Discussion of the Experiment

As it follows from the results of the experiment, non-stationary electric spinning, at least at the shown frequency and the field strength, lowers the average intensity of background radiation within the standard deviation. Another important detail of the results is an obvious reduction of the standard deviation for clockwise spinning. It's important to stress that for Poisson distribution spread of counts is calculated as a root square of the average. As the number of sampling increases, this value asymptotically approaches to the standard deviation. That is, a notable difference in standard deviation for the clockwise spinning, derived from 120 samplings, indicates that increased number of them will keep the same tendency.

Lowering the standard deviation implies increasing chances of appearance of several equal counts in row. This is confirmed by the table 2, where clockwise spinning results in appearing a number of the similar counts. This number considerably exceeds that calculated from a simple probability.

This is related to other noted behavior: The clockwise spinning breaks the symmetry of Poisson distribution. While absence of the spinning brings the distribution which is expected to be like Gaussian one as the number of samplings creases, the clockwise spinning breaks its normal shape. It has to be said that the counterclockwise spinning also shows this tendency, however less considerably.

It's interesting to note, that both passive and active torsion technology bring the tendency to reduce background gamma-radiation. Our results look pretty much alike those reported in [12]. The authors reported lowering the radiation from 20-12 uRm/h to 10-6 uRem/h. The shape-based sources reduced it from14 uRem/h to 8 uRem/h.

The observed results can be explained due to lowering fluctuations of vacuum. Being a virtual gyroscope, electric spinning influences a total angular momentum of vacuum, stabilizing its direction along the axis of the spinning within the frame of the conservation law.

The background gamma-radiation comes from space (i.e., cosmic rays) and from naturally occurring radioactive materials contained in the earth and in living things. Some portion of that is a result of vacuum fluctuations,

Fluctuations of physical vacuum can result in originating virtual quanta directly from the vacuum. Today's Physics already has the experience of realization of vacuum fluctuation on a macro-level: Casimir effect, predicted in1948 [18]. The experiment, considered by H. Casimir as a proof of statistical nature of Quantum Mechanics, was finally successfully accomplished in 1996 and demonstrated originating mechanical force of vacuum fluctuations. Unlike the Casimir effect with the passive plate- resonators, the experiment with a virtual gyroscope actively disturb physical vacuum.

Then, staying within frames of the conservation law, we can explain reduction of the background radiation as a direct implementation of Le Chatelier-Braun principle: when a constrain is applied to a dynamic system in equilibrium, a change takes place within the system, opposing the constrain and tending to restore the equilibrium. In our experiment, the system consists of the vacuum fluctuations and superimposed spinning field. Minimization of free energy requires anti parallel orientation of spin when the spinning field is applied. This is illustrated by the Fig.7 where allocation **a** is more thermodynamically profitable than **b**.

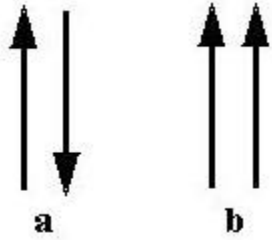

Fig.7
Anti parallel orientation of spin vectors is more thermodynamically profitable than the parallel one.

In the experiment, domination of the clockwise spinning can be attributed to predominant quanta having counterclockwise spins.

In the experiment, the angular momentum of fermions of the spinning electric field compensate action of the total spin of some number *a* of gamma-quanta, which are bosons by their nature.

$$\frac{\varepsilon\varepsilon_0 \int_V E^2(t)dV}{4c^2} r^2 \vec{\omega} = \hbar \left\langle \sqrt{s(s+1)} \right\rangle a \qquad (9)$$

Here, s=0,1,2..

Compensating total action of spin angular momentums can be considered as elimination of the particle, what explains reduction of background radiation in the spinning field.

On the other hand, the result of the experiment can be explained on a base of Heisenberg Uncertainty Principle

$$(\Delta p)(\Delta x) \geq \frac{\hbar}{2} \qquad (10)$$

We believe that the conservation of probability is applicable here. This is the requirement that the sum of the probabilities of finding a system in each of its possible states is constant. Indeed, before superposition of the spinning field and gamma-quanta, each the system had its own statistics. That is, the product $\Delta p \Delta x$ remains unchanged after the superposition. Electric spinning stabilizes a momentum of the quanta, lowering its uncertainty. As a result of the conservation of probability, the coordinate uncertainty $\Delta x$ increases. This means that the chances of meeting the gamma-quanta and the detector get lower.

On the other hand, electric spinning results in distortion of time. The Uncertainty Principle also can be shown as

$$(\Delta E)(\Delta t) \geq \frac{\hbar}{2} \qquad (11)$$

The superimposed modulated spinning definitely broadens the spectrum of the combined system, composed of the virtual gyroscope and gamma-quanta. That is, *ΔE* gets more. Then, the conservation of probability results in decreasing *Δt*, which can be considered as a time of interaction of the particle with the detector. However, decreasing uncertainty *Δt* could be a result of total retardation of time flowing in the context of space-time distortions as the rotating field is superimposed.

It's interesting to compare this corollary with the results of [12], where its authors measured distortions of time in the experiments with rotating mechanical objects. As it was cited above, they proved both retardation and acceleration of the time flow depending on the direction of the rotation. In [12], the most profound time-distortion was observed at the clockwise spinning. In our research, notable reduction of gamma-radiation also took place at the clockwise spinning.

Let's estimate the mass, angular momentum and energy of the virtual gyroscope. At E=100V/m , r =5E-2 m, and ω = 39E+3 rad/s we get the following:
m= 1.28E-28 kg; J= 6E-27 J*s; W= 1.2E-22 J. As we see, the spinning field has a mass, comparable to that of elemental particle. The total angular momentum of the gyroscope drastically exceeds that of the elemental spins of gamma-quanta and potentially can influence large number of them.

This energy can be compared to the differences energies of vacuum fluctuations in Casimir effect, estimated by him in [18].

$$\Delta E = \frac{\pi^2}{720} \frac{\hbar c}{L^3} A, \qquad (12)$$

where A and L are area and distance between the plates in Casimir–effect related experiment.

Calculation according to dimension of the cells brings ΔE=1.7E-26 J. When this result is compared with that of the field-gyroscope, E = 1.2E-22 J, it's apparent that the spinning field can control less powerful vacuum fluctuations.

**3.4. Role of Non-Stationary Rotation**

As it follows from formulas (4-6), modulating electric rotation as a manifestation of the non-stationary, widens a spectrum of the virtual gyroscope. Therefore, chances of interacting fermions of the virtual gyroscope with bosons of gamma-radiation increase. If the wider spectrum, the more interactions approach is a true, then application of frequency or phase modulations, having wide-band spectra is more effective in altering statistics of background gamma-radiation and time-related phenomena than the amplitude modulation.

## Conclusion

The experiments on influence of spinning electric field on background gamma-radiation revealed lowering intensity of the radiation in the spinning electric field.

The clockwise spinning causes more intensive influence on the radiation, resulting in breaking symmetry of Poisson distribution at large number of samplings.

Predominating role of the clockwise spinning is explained due to exceeding quantity of the gamma-particles having counterclockwise spinning in the same reference frame.

The lowering number of gamma-quanta and distortion of their statistics is explained due to stabilizing action of the field gyroscope as implementation of angular momentum conservation law for the combined system and thermodynamically profitable reorientation of initial vectors of the gamma-quanta spins.

Another explanation of the observed phenomenon is based on assumption of applicability of probability conservation here. Electric spinning reduces the uncertainty of the momentum and, following Heisenberg Uncertainty Principle, the uncertainty of coordinates of gamma-particle increases, reducing its chances of interacting with the detector. The same approach also leads to possibility of retardation of the time flow within the virtual gyroscope in the context of space-time distortion.

Comparison between energy of vacuum oscillations in Casimir effect, reduced to same size of the experimental cell, and the energy of the virtual gyroscope shows the possibility of the spinning field to control vacuum fluctuations.

Similarity of the gamma-radiation reducing effect produced by mechanical rotation and that of the spinning electric filed validates the conception of virtual gyroscope and stresses a special role of rotation itself, disregarding the matter of its carrier.

## Acknowledgement

The authors express their gratitude to engineer Igor Novikov, furnished Quartex counter for the experiments.

## References and links